# SIDE INFLUENCES ON THE OPERATION OF SPACE-BASED INTERFEROMETERS, AS INFERRED FROM LLR DATA

## Yu. V. Dumin

*Space Information Technology Center,
IZMIRAN, Russian Academy of Sciences
Troitsk, Moscow reg., 142190 Russia
E-mail: dumin@cityline.ru*

## HISTORY AND BASIC ACHIEVEMENTS IN LLR

Precise measurements of the Earth–Moon distance by using the ultrashort laser pulses – lunar laser ranging (LLR) – are carried out for over 30 years, after the installation of several retroreflectors on the lunar surface in the course of *Apollo* (USA) and *Lunakhod* (USSR) space missions [1]. The typical accuracy of these measurements was [1, 2]:
- $\sim 25$ cm in the early 1970's,
- $2$–$3$ cm in the late 1980's, and
- a few millimeters at the present time.

LLR works contributed significantly to astrometry, geodesy, geophysics, lunar planetology, and gravitational physics.

The most important results related to General Relativity are
(i) verification of the Strong Equivalence Principle with accuracy up to $\sim 10^{-13}$,
(ii) determination of the first post-Newtonian parameters in the gravitational field equations,
(iii) detection of the geodetic precession of the lunar orbit, and
(iv) imposing the observational constraints on time variations in the gravitational constant $G$ with accuracy $\sim 10^{-11}$ per year.

# AN UNRESOLVED PROBLEM: ANOMALOUS INCREASE IN THE LUNAR SEMIMAJOR AXIS

Despite the considerable advances listed above, there is a long-standing unresolved problem in the interpretation of LLR data– anomalous increase in the lunar semimajor axis [3].

In general, such increase is well-known and can be explained for the most part by tidal interaction between the Earth and Moon: Because of the relaxation processes, the tidal bulge is not perfectly symmetric about the Earth–Moon line but slightly shifted in the direction of Earth's rotation (see figure) [4]. As a result, there is a torque moment, which decelerates a proper rotation of the Earth and accelerates an orbital rotation of the Moon; so that the Earth–Moon distance increases.

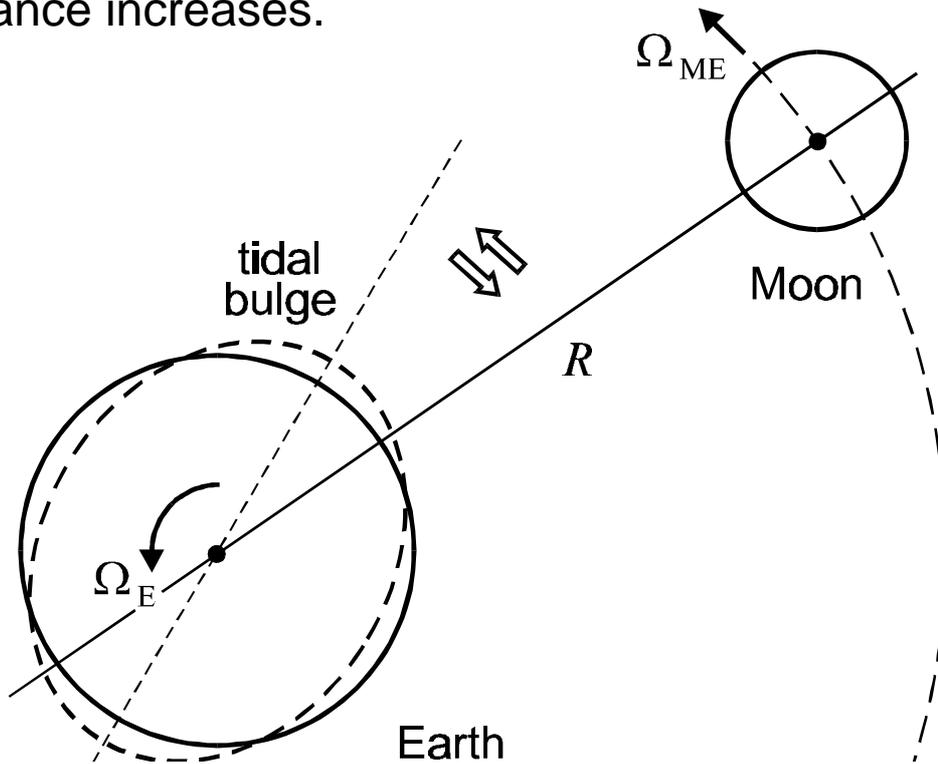

From the angular momentum conservation law

$$I_E \frac{d}{dt}\Omega_E + m_M \frac{d}{dt}\left(R^2 \Omega_{ME}\right) = 0$$

and the relation between the lunar orbital velocity and its distance from the Earth

$$\Omega_{ME} = G^{1/2} m_E^{1/2} R^{-3/2},$$

it can be easily found that the rate of increase in the lunar semi-major axis $\dot{R}$ is related to the rate of deceleration of the Earth's proper rotation $\dot{T}_E$ as

$$\dot{R} = k\,\dot{T}_E,$$

$$k = 4\pi\,G^{-1/2} I_E m_E^{-1/2} m_M^{-1} R^{1/2} T_E^{-2} = 1.81\cdot 10^5 \text{ cm/s},$$

where $\Omega_E$ and $T_E$ are the angular velocity and period of the proper rotation of the Earth, $\Omega_{ME}$ is the angular velocity of orbital rotation of the Moon about the Earth, $R$ is the distance between them (i.e. the semimajor axis), $I_E$ is the Earth's moment of inertia, $m_E$ and $m_M$ are the terrestrial and lunar masses, and $G$ is the gravitational constant.

According to telescopic observations of the Earth's rotation with respect to the distant objects, $\dot{T}_E^{(\text{tel})} = 1.4\cdot 10^{-5}$ s/yr; so that $\dot{R}^{(\text{tel})} = 2.53$ cm/yr. On the other hand, immediate LLR measurements give an appreciably greater value $\dot{R}^{(\text{LLR})} = 3.82$ cm/yr [1, 3].

## PROBABLE INTERPRETATIONS OF THE ANOMALY

The most of attempts to explain the above-mentioned anomaly $\Delta\dot{R} = \dot{R}^{(\text{LLR})} - \dot{R}^{(\text{tel})} = 1.29$ cm/yr were based on accounting for extra geophysical effects (e.g. secular changes in the Earth's moment of inertia $I_E$). Unfortunately, they did not lead to a satisfactory quantitative agreement with observations.

Yet another promising interpretation of the anomaly $\Delta\dot{R}$ is to attribute this discrepancy to the "residual" Hubble expansion in the local space environment [5]. As follows from the standard relation $\Delta\dot{R} = H_0^{(\text{loc})} R$, the "local" Hubble constant should be $H_0^{(\text{loc})} =$ $33 \pm 5$ (km/s)/Mpc (i.e. only about one-half its commonly-accepted value at the intergalactic scale).

In general, the small value of $H_0^{(\text{loc})}$ presented above is not surprising, because the local Hubble expansion should be formed only by some kinds of *unclumped* "dark matter" or "dark energy", uniformly distributed in the Universe (such as $\Lambda$-term, inflaton-like scalar field, and so on); while the other kinds of matter experienced gravitational instability, formed compact objects, and are no longer able to make their contributions to Hubble flow at the small scales.

Besides, apart from any theoretical arguments in favor of Hubble expansion in the local space environment, the recent observations [6] revealed that a quiescent Hubble flow begins at least from the distances $\sim 1$–$2$ Mpc, i.e., an order of magnitude less than was usually expected before. This fact was also interpreted as contribution of an unclumped dark matter to the formation of Hubble expansion [7].

## INFLUENCE ON THE OPERATION OF SPACE-BASED INTERFEROMETERS

By summarizing the above discussion, we can conclude that the nature of anomalous increase in the Earth–Moon distance is not quite clear yet. Nevertheless, this phenomenon may result in important consequences for the operation of space-based laser interferometers, designed for searching the gravitational waves (such as *LISA* [8] and *ASTROD* [9]).

First of all, if the local Hubble expansion really exists, it will produce an appreciable systematic variation in the interferometer arm length (with amplitude $\dot{L}/L \approx 1.1 \cdot 10^{-18}$ s$^{-1}$) and, therefore, should be inevitably taken into account in the processing of gravitational-wave data.

Second, since the space-based interferometers do not suffer from any geophysical uncertainties (such as an efficiency of the Earth's response to the tidal perturbations, secular variations in the Earth's moment of inertia, and so on), the nature of the anomaly under consideration can be unambiguously revealed. *Moreover, if the LLR anomaly is actually caused by the local Hubble expansion, the space interferometers will be a valuable tool for immediate cosmological studies.*

Finally, it should be emphasized that the requirements for measuring the anomalous secular variation in the interferometer arm length seem to be even less hard than for searching the gravitational-wave signals. *So, the respective experiments may be carried out at a preliminary stage of testing the drag-free satellites, before the full-scale deployment of the entire space interferometric system.*